\documentclass{iaus}
\usepackage{graphics}

  \checkfont{eurm10}
  \iffontfound
    \IfFileExists{upmath.sty}
      {\typeout{^^JFound AMS Euler Roman fonts on the system,
                   using the 'upmath' package.^^J}%
       \usepackage{upmath}}
      {\typeout{^^JFound AMS Euler Roman fonts on the system, but you
                   dont seem to have the}%
       \typeout{'upmath' package installed. iaus.cls can take advantage
                 of these fonts,^^Jif you use 'upmath' package.^^J}%
      }
  \else
  \fi

  \checkfont{msam10}
  \iffontfound
    \IfFileExists{amssymb.sty}
      {\typeout{^^JFound AMS Symbol fonts on the system, using the
                'amssymb' package.^^J}%
       \usepackage{amssymb}%

      }{}
  \fi

  \IfFileExists{amsbsy.sty}
    {\typeout{^^JFound the 'amsbsy' package on the system, using it.^^J}%
     \usepackage{amsbsy}}
    {}

\newsavebox{\astrutbox}
\sbox{\astrutbox}{\rule[-5pt]{0pt}{20pt}}

\title[Outskirts of Galaxy Clusters: intense life in the suburbs] 
{Observing the build--up of the colour-magnitude relation at redshift $\sim
  0.8$\thanks{Based on observations obtained at the ESO Very Large Telescope
  (VLT) as part of the Large Programme 166.A--0162 
  (the ESO Distant Cluster Survey).}}

\author[Gabriella De Lucia {\it et al.\/}]{G. De Lucia$^1$, B. M.
  Poggianti$^2$, A.  Arag\'on-Salamanca$^3$, D.  Clowe$^4$, C.  Halliday$^2$,
  P. Jablonka$^5$, B.  Milvang-Jensen$^6$, R.  Pell\'o$^7$, S. Poirier$^5$, G.
  Rudnick$^1$, R. Saglia$^6$, L.  Simard$^8$, S.~D.~M. White$^1$ \& \\
  the EDisCS collaboration}

\affiliation{$^1$MPA, Garching, Germany -- $^2$Osservatorio Astronomico di
  Padova, Italy -- $^3$University of Nottingham, UK -- $^4$Steward
  Observatory, Tucson, USA -- $^5$OPM, Paris, France -- $^6$MPE, Garching,
  Germany -- $^7$OMP Toulouse, France -- $^8$ HIA, Victoria, Canada}

\pubyear{2004}
\volume{195}
\pagerange{1--8}
\date{?? and in revised form ??}
\jname{Outskirts of Galaxy Clusters: intense life in the suburbs}
\editors{A. Diaferio, ed.}
\begin{document}

\maketitle

\begin{abstract}
  We analyse the rest--frame (U$-$V) colour--magnitude relation for $2$
  clusters at redshift $0.7$ and $0.8$, drawn from the ESO Distant Cluster
  Survey.  By comparing with the population of red galaxies in the Coma
  cluster, we show that the high redshift clusters exhibit a deficit of passive
  faint red galaxies.  Our results show that the red--sequence population
  cannot be explained in terms of a monolithic and synchronous formation
  scenario.  A large fraction of faint passive galaxies in clusters today has
  moved onto the red sequence relatively recently as a consequence of the fact
  that their star formation activity has come to an end at $z<0.8$.
\end{abstract}

\firstsection 
\section{Introduction}
\label{sec:intro}
Red cluster galaxies form a tight sequence in the colour--magnitude diagram
that, in nearby clusters, extends over a range of at least $5$--$6$ mag from
the Brightest Cluster Galaxy (BCG).  The existence of a tight colour--magnitude
relation (CMR) up to redshift $\sim 1$, and the evolution of its slope and its
zero--point as a function of redshift, are commonly interpreted as the result
of a single formation scenario in which cluster ellipticals constitute a
passively evolving population formed at high redshift ($z\gtrsim2$--$3$) in a
monolithic collapse (\cite[Kodama et al. 1998]{K98}).  In this model, the slope
of the CMR reflects metallicity differences and naturally arises taking into
account the effects of supernovae winds.  An alternative explanation has been
proposed by \cite{KC} and confirmed recently by \cite{DLKW}.  In this model,
elliptical galaxies form through mergers of disk systems and a CMR arises
because more massive ellipticals form by mergers of more massive, and hence
more metal rich, disk systems.

On the other hand, it is clear that red passive galaxies in distant clusters
constitute only a subset of the passive galaxy population in clusters today
(\cite[van Dokkum \& Franx 1996]{DF}).  Distant clusters contain significant
populations of galaxies with active star formation, that can evolve onto the
CMR after their star formation activity is terminated, possibly as a
consequence of their infall onto clusters (\cite[Smail et al. 1998; Poggianti
et al. 1999]{S98,P99}).

In this work we present the CMR for $2$ clusters at redshifts $0.7$ and $0.8$
from the ESO Distant Cluster Survey (EDisCS).

\section{The ESO Distant Cluster Survey}
\label{sec:ediscs}
EDisCS is an ESO Large Programme aimed at the study of cluster structure and
cluster galaxy evolution over a significant fraction of cosmic time.  It will
provide homogeneous photometry and spectroscopy for $10$ clusters at redshift
$\sim 0.4-0.5$ and $10$ clusters at redshift $\sim 0.7-0.8$.  

In this work we present results for $2$ clusters at redshift $0.7$
(cl$1054.4$--$1146$) and $0.8$ (cl$1216.8$--$1201$).  The clusters have been
imaged in V, R, and I with FORS2 at the ESO's Very Large Telescope.  The
average integration times used were $115$ m in the I--band, and $120$ m in the
V--band.  Multi--object spectroscopy was carried out using FORS2 on the VLT.
All data and details about their analysis will be presented in forthcoming
papers (White et al, Halliday et al., in preparation; see also Halliday et al.,
these proceedings).  Objects detection has been performed using SExtractor
(\cite[Bertin \& Arnouts 1996]{BA}) in `two--image' mode using the I--band
images as detection reference images.  In the following, we will use magnitude
and colours measured on the seeing--matched (to $0.8$ arcsec -- our `worst'
seeing) registered frames using a fixed circular aperture with $1$ arcsec
radius.  At the clusters redshifts, this aperture corresponds to a physical
radius of $7.13$--$7.50$ kpc (we use $\Omega_{\rm m} = 0.3$,
$\Omega_{\Lambda}=0.7$, and $H_0 = 70\,{\rm km}\,{\rm s}^{-1}\,{\rm
  Mpc}^{-1}$).

\vspace{0.5cm}
\section{Method and Results}
\label{sec:results}
We have computed photometric redshifts using two independent codes (see
\cite[Rudnick et al. 2003]{R03}) and we have rejected non--members using a
two--step procedure: (i) we use the full redshift probability distributions
from the codes to isolate objects with high probability to be at the cluster
redshift and (ii) we perform a statistical subtraction on the remaining objects
using the distribution on the CMR of all the objects at distance from the BCG
larger than $1$ Mpc.

\begin{figure}
\begin{center}
\resizebox{14cm}{!}{\includegraphics{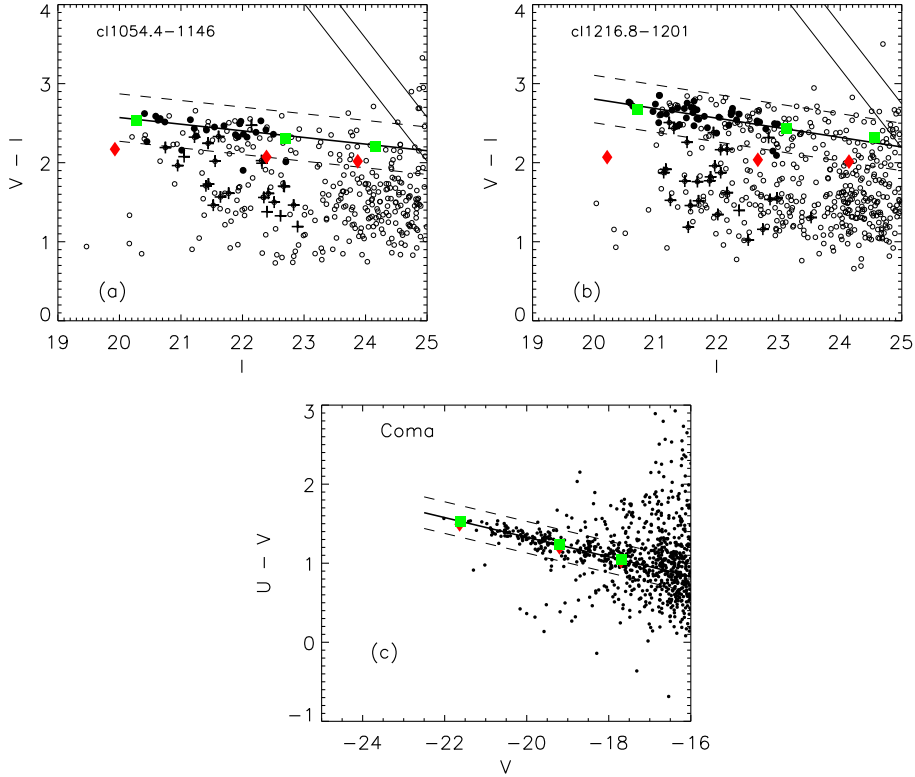}}
\caption{CM diagrams for the $2$ EDisCS clusters under investigation (panels a
  and b) and for the Coma cluster (panel c).  The solid thick line represents
  the best fit relation measured using a bi--weight estimator and only the
  spectroscopic members with absorption spectra.  The dashed lines correspond
  to $\pm 3\sigma$ from the best fit, where $\sigma$ is the dispersion of the
  objects used for the fit. Boxes and diamonds show the location of models with
  different star formation histories (see text for detail). [\emph{See online
    version of the proceeding book for a colour version of this figure}].}
\label{fig:cm}
\end{center}
\end{figure}

Fig.~\ref{fig:cm} (panels a and b) shows the V$-$I diagram for the $2$ clusters
used in this analysis.  At these redshifts, V$-$I approximatively samples the
rest--frame U$-$V colour, and is therefore very sensitive to any recent or
ongoing star formation episode.  A red sequence is clearly visible for each
cluster, together with a significant population of the blue galaxies known to
populate clusters at high redshift.  Thin solid lines correspond to $3$ and
$5\sigma$ detection limits in the V--band.  Empty symbols represent galaxies
retained as cluster members; filled circles and crosses are spectroscopically
confirmed members with absorption and emission--line spectra respectively.
Panel (c) of Fig.~\ref{fig:cm} shows the U$-$V CMR for the Coma cluster
including all galaxies in the catalogue by \cite{TCB}.  Here we use magnitudes
and colours in a $25.2$ arcsec diameter aperture that, at the redshift of Coma,
corresponds to a physical size of $11.71$ kpc and therefore quite closely
approximates our $\sim 14$ kpc diameter aperture at $z =0.8$.  We have assumed
a distance modulus of $35.16$ and converted observed colours to rest--frame
colours using tabulated K--corrections (\cite[Poggianti 1997]{P97}). 

The green boxes in Fig.~\ref{fig:cm} show the location of a model with a single
burst at $z=3$, and the red diamonds show the location of a model with a $1$
Gyr exponentially declining SFR starting at $z=3$, calculated with the code by
\cite{BC}.  Three different metallicities are shown: $0.02$, $0.008$ and
$0.004$ (from brighter to fainter).  The single burst model provides a
remarkably good fit to the CMR of the high redshift clusters, confirming that
the location of the CM sequence observed in distant clusters requires high
redshifts of formation, and that the slope is consistent with a correlation
between galaxy metal content and luminosity.  

\begin{figure}
\begin{center}
\vspace{0.5cm}
\resizebox{14.5cm}{!}{\includegraphics{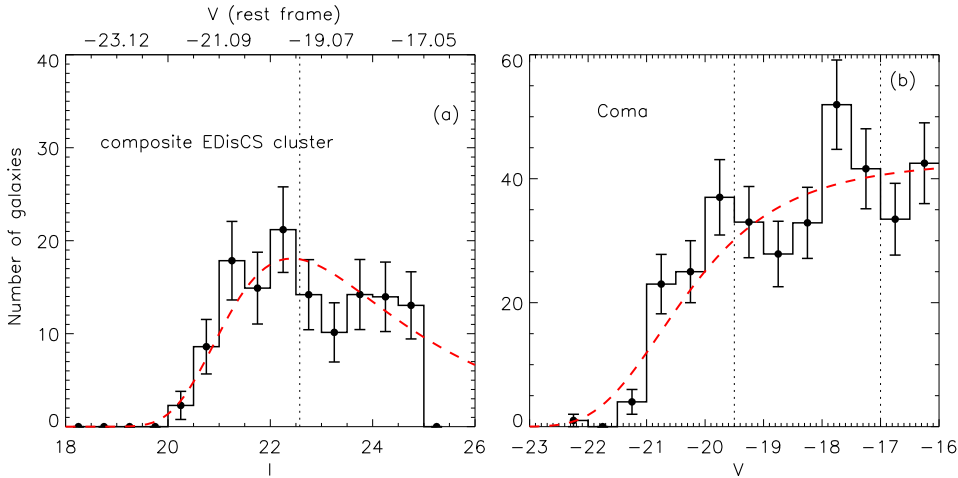}}
\caption{LF of red-sequence galaxies for a composite cluster at
  redshift $0.75$ (panel a) and for the Coma cluster (panel b).  Poisson error
  bars are shown.  The dashed lines represent a fit with a Schechter function.
  The scale on the top of the left panel has been obtained from the observed
  I--band magnitudes after correcting for passive evolution.  Dotted vertical
  lines show the limits used to define the luminous--to--faint ratio (see
  text for detail).}
\label{fig:histo}
\end{center}
\end{figure}

However, Fig.~\ref{fig:cm} also shows that the CMR in our clusters is
strikingly `empty' at magnitudes fainter than $\sim 23$.  In order to quantify
this effect, we compute the luminosity function (LF) of red sequence galaxies.
Fig.~\ref{fig:histo} shows the numbers of \emph{bona fide} cluster members
within $\sim 3\sigma$ from the best fit relation as a function of galaxy
magnitudes.  For the EDisCS clusters, we combine the results for the $2$
clusters correcting colours and magnitudes to a common redshift $=0.75$ and
average over $100$ realizations of the statistical subtraction for each
cluster.  For Coma, we have corrected the number of red galaxies in each
magnitude bin for background and foreground contamination using a redshift
catalogue kindly provided by Matthew Colless and the same procedure as in
\cite{M03}.

The histogram of the EDisCS clusters shows a remarkable deficiency of $\lesssim
0.1$L$_*$ red sequence galaxies.  In contrast, the number of passive red
galaxies in Coma tends to increase going to fainter magnitudes, in agreement
with the luminosity function of early spectral types in clusters in the 2dF
Galaxy Redshift Survey (\cite[De Propris et al. 2003]{DP03}).  A Schechter fit
to the histograms shown in Fig.~\ref{fig:histo} yields a faint--end slope of
$-0.82$ for the EDisCS clusters and $-1$ for Coma, but the error bars are quite
large and therefore the results of the fit are just indicative.  

If we arbitrarily classify as `luminous' galaxies those brighter than $M_{\rm
  V}= -19.5$ (this corresponds to an observed I--band magnitude of $22.57$ at
redshift $0.75$), and as `faint' those galaxies that are fainter than this
magnitude and brighter than $-17$ (this corresponds to the magnitude limit for
our high redshift clusters), we obtain a luminous--to--faint ratio for the Coma
cluster $0.48 \pm 0.06$, where the error has been estimated assuming a Poisson
statistics.  The corresponding value for the composite EDisCS cluster is
$0.99\pm0.17$.  This ratio therefore differs between the composite high
redshift cluster and the Coma cluster at about a $3\sigma$ level.
\section{Discussion and Conclusions}
\label{sec:discussion}

The results presented are robust both against the technique adopted for
removing non--cluster members and against the photometric errors.  The red
galaxy deficit is detected also when rejecting non--members using a purely
statistical subtraction or a more stringent criterion for membership based
solely on photometric redshifts.  In fact, a deficit is evident also in the
{\it full} photometric catalogue, when no field correction is attempted.
Photometric errors in the EDisCS catalogue are comparable to the errors for the
Coma data, therefore the differences observed in the distributions of
Fig.~\ref{fig:histo} cannot be a spurious result arising from photometric
errors.

A decline in the number of red sequence members at faint magnitudes was first
observed in clusters at $z=0.25$ by \cite{S98}.  Evidence for a `truncation'
of the CM sequence has been noticed in a cluster at $z=1.2$ by \cite{K00}, who
suggested that faint early--type galaxies might not have been in place until $z
\sim 1.2$.  In more recent work \cite{K04} have come to the same conclusion
using deep wide-field optical imaging data of the Subaru/XMM-Newton Deep
Survey. 

The CMR of the $2$ EDisCS clusters at $z\sim0.8$ also shows a deficit of red,
relatively faint galaxies.  Our investigation shows that the evolution of the
red--sequence population at these redshifts cannot be explained as the result
of a monolithic and synchronous formation scenario.  A large fraction of the
passive $\lesssim 0.1$L$_*$ galaxies must have moved onto the CMR at redshifts
lower than $0.8$ as a consequence of the fact that their star formation
activity has come to an end.  Our analysis enlightens the importance of
studying the evolution of the cluster population as a whole, trying to
understand how galaxies are accreted from the field, how the dense cluster
environment affects their star formation rate, and how these galaxies fade and
move on to the red sequence.  Only this kind of analysis can unveil the full
evolutionary paths of galaxies that lie on the red--sequence today and give
strong constraints on the relative importance of star formation and metallicity
in establishing the the observed red--sequence.  We plan to investigate this in
more detail in future work.

\begin{acknowledgments}
We thank F. La Barbera, S. Zaroubi, I. Smail and M. Colless.
G.~D.~L. thanks the Alexander von Humboldt Foundation, the Federal Ministry of
Education and Research, and the Programme for Investment in the Future (ZIP) of
the German Government for financial support.
\end{acknowledgments}

\end{document}